\documentclass[epj,nopacs]{svjour}
\usepackage{graphics}
\usepackage{bm}        
\usepackage{amssymb}   
\usepackage[utf8]{inputenc}
\usepackage{mathrsfs}
\usepackage{amsmath}
\usepackage{hyperref}
\usepackage{textcomp}
\usepackage{color}
\usepackage{epsfig}
\usepackage{multirow}
\usepackage{textcomp,color}
\usepackage{graphicx,subfigure}
\usepackage{multicol}
\usepackage[normalem]{ulem}
\usepackage[english]{babel}

\begin{document}

\title{Scattering of the double sine-Gordon kinks}
\author{Vakhid A. Gani\inst{1,2}\and Aliakbar Moradi Marjaneh\inst{3}\and Alidad Askari\inst{4}\and Ekaterina Belendryasova\inst{1}\and Danial Saadatmand\inst{5}
\mail{moradimarjaneh@gmail.com (A.M.M., corresponding author) and vagani@mephi.ru (V.A.G.).}
}                     
%
%
\institute{National Research Nuclear University MEPhI (Moscow Engineering Physics Institute), 115409 Moscow, Russia \and Theory Department, National Research Center Kurchatov Institute, Institute for Theoretical and Experimental Physics, 117218 Moscow, Russia \and Young Researchers and Elite Club, Quchan Branch, Islamic Azad university, Quchan, Iran \and Department of Physics, Faculty of Science, University of Hormozgan, P.O.Box 3995, Bandar Abbas, Iran \and Department of Physics, University of Sistan and Baluchestan, Zahedan, Iran}
\date{Received: date / Revised version: date}
%
\abstract{
We study the scattering of kink and antikink of the double sine-Gordon model. There is a critical value of the initial velocity $v_\mathrm{cr}$ of the colliding kinks, which separates different regimes of the collision. At $v_\mathrm{in}>v_\mathrm{cr}$ we observe kinks reflection, while at $v_\mathrm{in}<v_\mathrm{cr}$ their interaction is complicated with capture and escape windows. We obtain the dependence of $v_\mathrm{cr}$ on the parameter of the model. This dependence possesses a series of local maxima, which has not been reported by other authors. At some initial velocities below the critical value we observe a new phenomenon --- the escape of two oscillons in the final state. Besides that, at $v_\mathrm{in}<v_\mathrm{cr}$ we found the initial kinks' velocities at which the oscillons do not escape, and the final configuration looks like a bound state of two oscillons.
\PACS{
      {PACS-key}{discribing text of that key}   \and
      {PACS-key}{discribing text of that key}
     } 
} 
\maketitle

\section{Introduction}
\label{sec:Introduction}


The $(1+1)$-dimensional field-theoretical models possessing the topologically non-trivial solutions --- kinks --- are of special interest for modern physics. They arise in a vast variety of models in quantum and classical field theory, high energy physics, cosmology, condensed matter physics, and so on, \cite{Rajaraman.book.1982}--\cite{Vachaspati.book.2006}.
Firstly, the $(1+1)$-dimensional models can be investigated analytically and numerically much easier than $(2+1)$ or $(3+1)$-dimensional. Because of that, some general properties of topological defects can be studied within the $(1+1)$-dimensional setups. Secondly, many physical systems can be effectively described by the one-dimensional structures. For example, a plane domain wall --- a wall, which separates regions with different vacuum states --- in the direction orthogonal to it, presents a kink. Surely, the topological defects arise and in more complex models with two, three or more fields. For example, in \cite{Bazeia.PLA.2013}--\cite{Alonso.2017.2}
the kink-like structures were studied in models with two interacting real scalar fields, for further information see also \cite{Lensky.JETP.2001.eng}--\cite{Ashcroft.JPA.2016}.

Kink-antikink collisions, as well as interactions of kinks with impurities (spatial inhomogeneities), are of growing interest since 1970s \cite{Kudryavtsev.UFN.1997.eng,Kudryavtsev.UFN.1997.rus}. Nevertheless, today it is a very fast developing area of research. For investigating of the kink-antikink interactions various approximate methods are widely used. Among them, the collective coordinate approximation \cite{GaKuLi}--\cite{GaKu.SuSy.2001.rus}.
Withing this method a real field system ``kink+antikink'' is approximately described as a system with one or several degrees of freedom. For instance, the kink-antikink separation can be used as the only (translational) degree of freedom. The more complicated modifications of the method have also been elaborated, which include other degrees of freedom (in particular, vibrational), see, {\it e.g.}, \cite{Weigel.cc.2014,Weigel.cc.2016,Demirkaya.cc.2017}.

Another approximate method for investigating the kinks interactions is Manton's method \cite[Ch.~5]{Manton.book.2004}, \cite{perring62}--\cite{KKS.PRE.2004}.
This method is based on using of the kinks asymptotics, it enables to estimate the force between the kink and the antikink at large separations.

On the other hand, recently the numerical simulation has become a powerful tool for studying the dynamics of the one-dimensional field systems. Using various numerical methods, many important results were obtained. In particular, the resonance phenomena --- escape windows and quasi-resonances --- have been found and investigated in the kinks' scattering \cite{GaKuLi}, \cite{Gani.PRE.1999}--\cite{Saadatmand.PRD.2015}.
Many important results have been obtained for the models with polynomial potentials of fourth, sixth, eighth, and higher degree self-interaction \cite{GaKuLi}, \cite{Dorey.PRL.2011}--\cite{Moradi.JHEP.2017}, \cite{Belendryasova.arXiv.2017,Belendryasova.conf.2017}, \cite{lohe}--\cite{Snelson.arXiv.2016}.
One should note interesting results on the long-range interaction between kink and antikink \cite{Radomskiy,Belendryasova.arXiv.2017,Belendryasova.conf.2017}, \cite{Guerrero.PRE.1997}--\cite{Gomes.PRD.2012}.
The models with non-polynomial potentials are also widely discussed in the literature, for example, the modified sine-Gordon \cite{Peyrard.msG.1983}, the multi-frequency sine-Gordon \cite{Delfino.NPB.1998}, the double sine-Gordon \cite{Gani.PRE.1999}, \cite{Campbell.1986}--\cite{Malomed.PLA.1989},
and a number of models, which can be obtained by using the deformation procedure \cite{Bazeia.arXiv.2017.sinh,Bazeia.arXiv.2017.sinh.conf,Bazeia.PRD.2006,Bazeia.PRD.2002,Bazeia.PRD.2004}.

The impressive progress is achieved in the investigation of domain walls, bubbles, vortices, strings, \cite{Campanelli.IJMPD.2004}--\cite{Gani.YaF.2001.rus},
as well as the embedded topological defects, {\it e.g.}, a Q-lump on a domain wall, a skyrmion on a domain wall, etc.~\cite{nitta1}--\cite{Bazeia.fermion.2017}.
Besides that, we have to mention various configurations of the type of Q-balls \cite{Schweitzer.PRD.2012.1}--\cite{Dzhunushaliev.PRD.2016}.
Topologically non-trivial field configurations could also lead to a variety of phenomena in the early Universe \cite{GaKiRu,GaKiRu.conf}.

In this paper we study the kink-antikink collisions within the double sine-Gordon model \cite{Gani.PRE.1999}, \cite{Campbell.dsG.1986}--\cite{Malomed.PLA.1989}.
There is a critical value of the initial velocity of the colliding kinks, $v_\mathrm{cr}^{}$, which separates different regimes of the collision. At $v_\mathrm{in}^{}>v_\mathrm{cr}^{}$ the kinks pass through each other and escape to infinities, while at $v_\mathrm{in}^{}<v_\mathrm{cr}^{}$ the kinks' capture and a complex picture of the so-called escape windows are observed, see, {\it e.g.}, \cite{Kudryavtsev.UFN.1997.eng,Kudryavtsev.UFN.1997.rus,Campbell.dsG.1986}. We performed a detailed study of the kink-antikink scattering at various values of the model parameter $R$. We have found a series of local maxima of the dependence of $v_\mathrm{cr}^{}$ on $R$, which has not been reported up to now. Besides that, at some initial velocities of the colliding kinks we observed final configuration of the type of two oscillons, which form a bound state or could escape to spatial infinities with some final velocities.

Our paper is organized as follows. In section \ref{sec:Topological_defects} we give some general information about the $(1+1)$-dimensional models with one real scalar field. Section \ref{sec:DSG_model} introduces the double sine-Gordon model, describes its potential, kinks, and their main properties. In section \ref{sec:Scattering} we study the scattering of the kink and antikink. In this section we present our main results related to the kink-antikink collisions. Finally, we summarize and formulate prospects for future works in section \ref{sec:Conclusion}.

\section{Topological defects in $(1+1)$ dimensions}
\label{sec:Topological_defects}

Consider a field-theoretical model in $(1+1)$-dimensional space-time with a real scalar field $\phi(x,t)$. The dynamics of the field $\phi$ is described by the Lagrangian density
\begin{equation}\label{eq:Largangian}
\mathcal{L} = \frac{1}{2}\left(\frac{\partial\phi}{\partial t}\right)^2 - \frac{1}{2}\left(\frac{\partial\phi}{\partial x}\right)^2 - V(\phi),
\end{equation}
where the potential $V(\phi)$ defines self-interaction of the field $\phi$. We assume that the potential is a non-negative function of $\phi$, which has a set of minima 
\begin{equation}
\mathcal{V}=\left\{\phi_1^{\mathrm{(vac)}}, \phi_2^{\mathrm{(vac)}}, \phi_3^{\mathrm{(vac)}},\dots\right\},
\end{equation}
which is a vacuum manifold of the model, and $V(\phi)=0$ for all $\phi\in\mathcal{V}$. The energy functional corresponding to the Lagrangian \eqref{eq:Largangian} is
\begin{equation}\label{eq:energy}
E[\phi] = \int\limits_{-\infty}^{\infty}\left[\frac{1}{2}\left(\frac{\partial\phi}{\partial t}\right)^2 + \frac{1}{2}\left(\frac{\partial\phi}{\partial x}\right)^2 + V(\phi)\right]dx.
\end{equation}
The Lagrangian \eqref{eq:Largangian} yields the equation of motion for the field $\phi(x,t)$:
\begin{equation}\label{eq:eqmo}
\frac{\partial^2\phi}{\partial t^2} - \frac{\partial^2\phi}{\partial x^2} + \frac{dV}{d\phi} = 0.
\end{equation}
In the static case $\phi=\phi(x)$, $\displaystyle\frac{\partial\phi}{\partial t}=0$, and we obtain
\begin{equation}\label{eq:eqmo_static}
\frac{d^2\phi}{dx^2} = \frac{dV}{d\phi}.
\end{equation}
This equation can be reduced to the first order ordinary differential equation
\begin{equation}\label{eq:bps}
\frac{d\phi}{dx} = \pm\sqrt{2V(\phi)}.
\end{equation}
In order for the energy of the static configuration to be finite, it is necessary that
\begin{equation}\label{eq:minus_infty}
\phi(-\infty) = \lim_{x \to -\infty} \phi (x) = \phi_i^{\mathrm{(vac)}}
\end{equation}
and
\begin{equation}\label{eq:plus_infty}
\phi(+\infty) = \lim_{x \to +\infty} \phi (x) = \phi_j^{\mathrm{(vac)}},
\end{equation}
where $\phi_i^{\mathrm{(vac)}},\phi_j^{\mathrm{(vac)}}\in\mathcal{V}$. If these two equalities hold, the second and the third terms in square brackets in \eqref{eq:energy} fall off at $x\to\pm\infty$ (the first term turns to zero for all static configurations), and the integral in \eqref{eq:energy} can be convergent.

If the vacuum manifold $\mathcal{V}$ consists of more than one point, i.e\ the potential $V(\phi)$ possesses two or more degenerate minima, the set of all static configurations with finite energy can be split into disjoint equivalence classes (or topological sectors) according to the asymptotic behaviour of the configuration at $x\to\pm\infty$. Configurations with $\phi_i^{\mathrm{(vac)}}\neq\phi_j^{\mathrm{(vac)}}$ in eqs.~\eqref{eq:minus_infty} and \eqref{eq:plus_infty} are called topological, while those with $\phi_i^{\mathrm{(vac)}}=\phi_j^{\mathrm{(vac)}}$ --- non-topological. A configuration belonging to one equivalence class (topological sector) can not be transformed into a configuration from another class (topological sector) through a continuous deformation, that is via a sequence of configurations with finite energies.

To describe the topological properties of the configurations, one can introduce a conserved topological current, {\it e.g.},
\begin{equation}\label{eq:top_current}
j_\mathrm{top}^\mu = \frac{1}{2}\varepsilon^{\mu\nu}\partial_\nu\phi,
\end{equation}
here $\varepsilon^{\mu\nu}$ stands for the Levi-Civita symbol, the indices $\mu$ and $\nu$ take values 0 and 1 for a $(1+1)$-dimensional configuration, and $\partial_0\phi\equiv\displaystyle\frac{\partial\phi}{\partial t}$, $\partial_1\phi\equiv\displaystyle\frac{\partial\phi}{\partial x}$. The corresponding topological charge does not depend on the behaviour of the field at finite $x$,
\begin{equation}\label{eq:top_charge}
Q_\mathrm{top} = \int\limits_{-\infty}^{\infty}j_\mathrm{top}^0dx = \frac{1}{2}\left[ \phi(+\infty)-\phi(-\infty)\right].
\end{equation}
The value of $Q_\mathrm{top}$ is determined only by the asymptotics \eqref{eq:minus_infty}, \eqref{eq:plus_infty} of the field. The topological charge \eqref{eq:top_charge} is conserved during the evolution of the configuration. Nevertheless, configurations from different topological sectors may have the same topological charge. At the same time, configurations with different topological charges necessarily belong to different topological sectors.

Further, for the non-negative potential $V(\phi)$ we can introduce the superpotential --- a smooth (continuously differentiable) function $W(\phi)$ of the field $\phi$:
\begin{equation}\label{eq:dwdfi}
V(\phi) = \frac{1}{2}\left(\frac{dW}{d\phi}\right)^2.
\end{equation}
Using this representation of the potential, the energy of a static configuration can be written as
\begin{equation}\label{eq:static_energy_with_bps}
E = E_\mathrm{BPS}^{} + \frac{1}{2}\int_{-\infty}^{\infty}\left(\frac{d\phi}{dx}\pm\frac{dW}{d\phi}\right)^2dx,
\end{equation}
where
\begin{equation}\label{eq:static_energy_bps}
E_\mathrm{BPS}^{} = \big|W[\phi(+\infty)]-W[\phi(-\infty)]\big|.
\end{equation}
Here the subscript ``BPS'' stands for Bogomolny, Prasad, and Sommerfield \cite{BPS1.eng,BPS1.rus,BPS2}. From eq.~\eqref{eq:static_energy_with_bps} one can see that, firstly, the energy of any static configuration is bounded from below by $E_\mathrm{BPS}^{}$,
\begin{equation}\label{eq:bound_BPS}
E \ge E_\mathrm{BPS}^{},
\end{equation}
and, secondly, the static configuration, which satisfies the equation
\begin{equation}\label{eq:bps_with_superpotential}
\displaystyle\frac{d\phi}{dx} = \pm\frac{dW}{d\phi},
\end{equation}
saturates the inequality \eqref{eq:bound_BPS}, i.e\ has the minimal energy \eqref{eq:static_energy_bps} among all the configurations within a given topological sector. The solutions of eq.~\eqref{eq:bps_with_superpotential} are called BPS (or BPS saturated) configurations. Note that eq.~\eqref{eq:bps_with_superpotential} coincides with eq.~\eqref{eq:bps}.

A kink is a BPS saturated topological solution $\phi_\mathrm{k}^{}(x)$ of eq.~\eqref{eq:bps}, which connects two neighboring vacua of the model, {\it i.e.} for the kink solution the values $\phi_i^{\mathrm{(vac)}}$ and $\phi_j^{\mathrm{(vac)}}$ in \eqref{eq:minus_infty}, \eqref{eq:plus_infty} are adjacent minima of the potential $V(\phi)$. Below we use the terms ``kink'' and ``antikink'' for solutions with $\phi_j^{\mathrm{(vac)}}>\phi_i^{\mathrm{(vac)}}$ and $\phi_j^{\mathrm{(vac)}}<\phi_i^{\mathrm{(vac)}}$, respectively. Nevertheless, in some cases we use ``kink'' for both solutions, just to be brief.

\section{The double sine-Gordon model}
\label{sec:DSG_model}

Consider the double sine-Gordon (DSG) model. The potential of the DSG model can be written in several different forms. Below in this section we briefly recall two of them, and after that we give a detailed introduction to the properties of the DSG model employed by us.

\subsection{The $\eta$-parameterized potential}

Recall that, {\it e.g.}, in papers \cite{Gani.PRE.1999,Campbell.dsG.1986} the following parameterization has been used:
\begin{equation}\label{eq:potential_eta}
V_\eta^{}(\phi) = \frac{4}{1+4|\eta|}\left(\eta\:(1-\cos\phi)+1+\cos\frac{\phi}{2}\right),
\end{equation}
where $\eta$ is a real parameter, $-\infty<\eta<+\infty$. It is easy to see that
\begin{equation}
V_\eta^{}(\phi) =
\begin{cases}
\cos\phi - 1\quad \mbox{for}\quad \eta\to-\infty,\\
1 - \cos\phi\quad \mbox{for}\quad \eta\to+\infty,\\
4\left(1+\cos\displaystyle\frac{\phi}{2}\right)\quad \mbox{for}\quad \eta=0,
\end{cases}
\end{equation}
{\it i.e.} the potential $V(\phi)$ reduces to a sine-Gordon form for the field $\phi$ at $\eta\to\pm\infty$, and for the field $\phi/2$ at $\eta=0$.

The shape of the potential \eqref{eq:potential_eta} crucially depends on the parameter $\eta$. Following \cite{Campbell.dsG.1986}, we can split all values of $\eta$ into four regions: $\eta<-\displaystyle\frac{1}{4}$, $-\displaystyle\frac{1}{4}<\eta<0$, $0<\eta<\displaystyle\frac{1}{4}$, and $\eta>\displaystyle\frac{1}{4}$.

1. At $\eta<-\displaystyle\frac{1}{4}$ the potential \eqref{eq:potential_eta} has two distinct types of minima, $\phi_n^{\mathrm{(vac)}}=4\pi n+\arccos\displaystyle\frac{1}{4\eta}$ and $\phi_m^{\mathrm{(vac)}}=4\pi m-\arccos\displaystyle\frac{1}{4\eta}$, $n,m=0,\pm 1,\pm 2,\dots$, degenerate in energy, $V(\phi_n^{\mathrm{(vac)}})=V(\phi_m^{\mathrm{(vac)}})=0$, which are separated by inequivalent barriers.

2. At $-\displaystyle\frac{1}{4}<\eta<0$ the potential \eqref{eq:potential_eta} has a single type of minima at $\phi_n^{\mathrm{(vac)}}=(2n+1)2\pi$ with $V(\phi_n^{\mathrm{(vac)}})=0$.

3. At $0<\eta<\displaystyle\frac{1}{4}$ the potential \eqref{eq:potential_eta} is structurally similar to the previous region, with the same set of minima.

4. At $\eta>\displaystyle\frac{1}{4}$ the minima of the potential \eqref{eq:potential_eta} are $\phi_n^{}=4n\pi$ with $V(\phi_n^{})=\displaystyle\frac{8}{1+4|\eta|}$, and $\phi_m^{\mathrm{(vac)}}=(2m+1)2\pi$ with $V(\phi_m^{\mathrm{(vac)}})=0$, $n,m=0,\pm 1,\pm 2,\dots$.

This our paper deals with the DSG model with the positive values of $\eta$. In this case, as we show below, it is convenient to introduce another positive parameter.

\subsection{The $R$-parameterized potential}

\begin{figure}[h]
\centering
\includegraphics[width=0.45\textwidth]{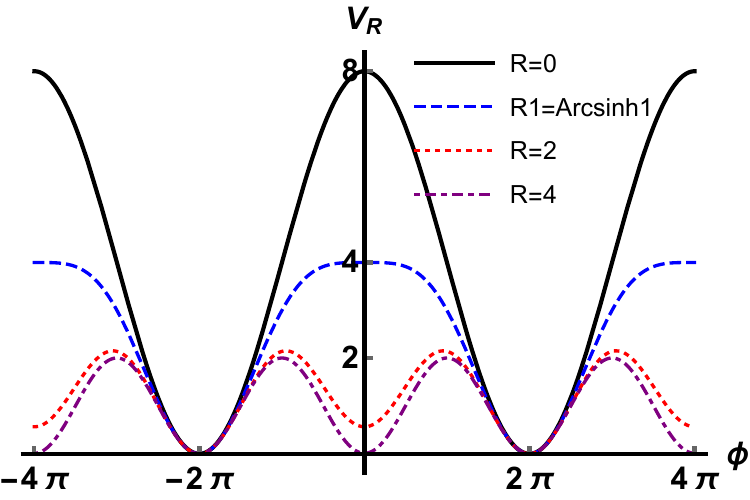} 
\caption{The potential \eqref{eq:potential_R} as a function of $\phi$ for various $R$.} 
\label{fig:PotentialPhi}
\end{figure}

For positive values of $\eta$, it is convenient to introduce another positive parameter, $R$, such that
\begin{equation}
\eta = \frac{1}{4}\sinh^2 R.
\end{equation}
In terms of the parameter $R$ the potential \eqref{eq:potential_eta} of the DSG model reads:
\begin{equation}\label{eq:potential_R}
V_\mathrm{R}^{}(\phi) = \tanh^2R\:(1-\cos\phi) + \frac{4}{\cosh^2R}\left(1+\cos\frac{\phi}{2}\right).
\end{equation}
Depending on the parameter $R$ the shape of the potential \eqref{eq:potential_R} looks different, see fig.~ \ref{fig:PotentialPhi}.
At $R=0$ we have a sine-Gordon potential for the field $\phi/2$, while at $R\to+\infty$ the potential \eqref{eq:potential_R} reduces to a sine-Gordon form for the field $\phi$:
\begin{equation}\label{eq:potential_R_limits}
V_\mathrm{R}^{}(\phi) =
\begin{cases}
4\left(1+\cos\displaystyle\frac{\phi}{2}\right)\quad \mbox{for}\quad R=0,\\
4-\left(1-\cos\displaystyle\frac{\phi}{2}\right)^2\quad \mbox{for}\quad R=\mbox{arcsinh}\:1,\\
1 - \cos\phi\quad \mbox{for}\quad R\to+\infty.
\end{cases}
\end{equation}

\subsection{The double sine-Gordon kinks}

\begin{figure*}[h!]
\begin{center}
  \centering
  \subfigure[Kinks ]{\includegraphics[width=0.45
 \textwidth]{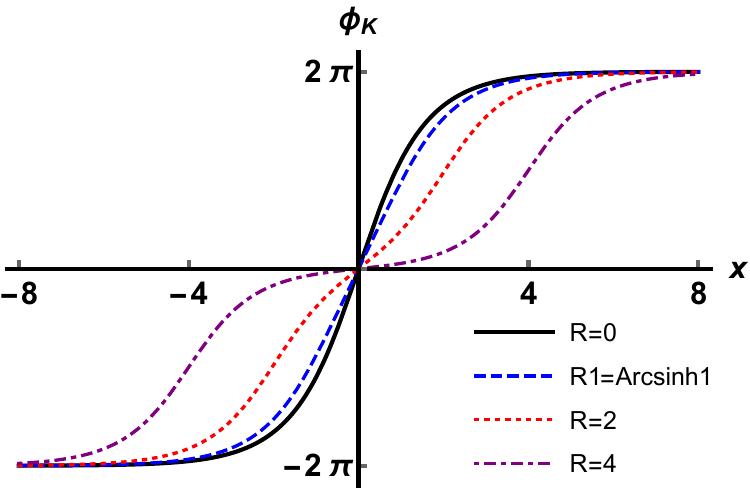}\label{fig:kinkSolutions}}
  \subfigure[Antikinks ]
{\includegraphics[width=0.45\textwidth]{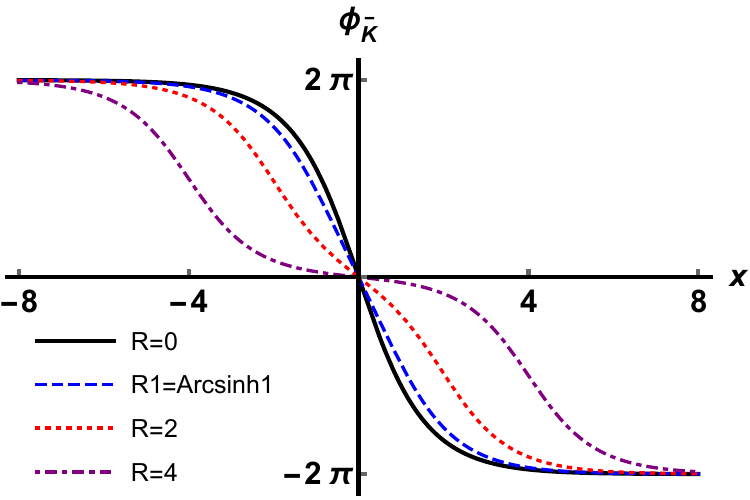}\label{fig:antikinkSolutions}}
  \\
  \caption{Kinks and antikinks \eqref{eq:DSG_kinks_1} for $n=0$ and various $R$'s.}
  \label{fig:SolitonSolutions}
\end{center}
\end{figure*}

\begin{figure}[h]
\centering
\includegraphics[width=0.45\textwidth]{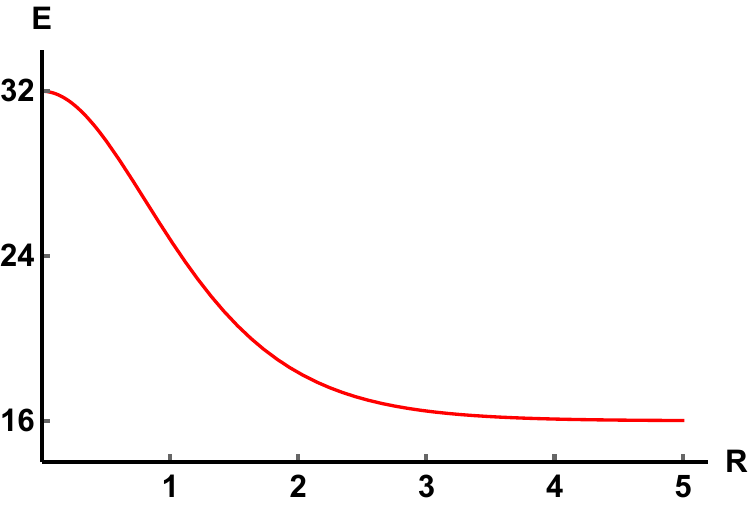} 
\caption{The energy of the static DSG kink (antikink) as a function of the parameter $R$.} 
\label{fig:kink_energy_vs_R}
\end{figure}

In terms of the parameter $R$ the static kink ($+$) and antikink ($-$) solutions can be written in a simple form,
\begin{equation}\label{eq:DSG_kinks_1}
\phi_{\mathrm{k}(\mathrm{\bar k})}(x) = 4\pi n \pm 4\arctan\frac{\sinh x}{\cosh R}.
\end{equation}
The DSG kink (antikink) can also be expressed as a superposition of two sine-Gordon solitons,
\begin{equation}\label{eq:DSG_kinks_2}
\phi_{\mathrm{k}(\mathrm{\bar k})}^{}(x) = 4\pi n \pm \left[\phi_\mathrm{SGK}^{}(x+R)-\phi_\mathrm{SGK}^{}(R-x)\right],
\end{equation}
or
\begin{equation}\label{eq:DSG_kink}
\phi_\mathrm{k}^{}(x) = 2\pi(2n-1) + \left[\phi_\mathrm{SGK}^{}(x+R)+\phi_\mathrm{SGK}^{}(x-R)\right]
\end{equation}
and
\begin{equation}\label{eq:DSG_antikink}
\phi_\mathrm{\bar k}^{}(x) = 2\pi(2n+1) - \left[\phi_\mathrm{SGK}^{}(x+R)+\phi_\mathrm{SGK}^{}(x-R)\right],
\end{equation}
where $\phi_\mathrm{SGK}^{}(x)=4\arctan\exp (x)$ is the sine-Gordon soliton. According to eqs.~\eqref{eq:DSG_kinks_2}--\eqref{eq:DSG_antikink}, the DSG kink can be viewed as the superposition of two sine-Gordon solitons, which are separated by the distance $2R$ and centered at $x=\pm R$, see fig.~\ref{fig:SolitonSolutions}.
The energy of the static DSG kink (antikink) is a function of the parameter $R$,
\begin{equation}\label{eq:DSG_kink_energy}
E(R) = 16\left(1+\frac{2R}{\sinh 2R}\right),
\end{equation}
this dependence is shown in fig.~\ref{fig:kink_energy_vs_R}.
Below we study the collisions of the DSG kinks. In such processes the kink's internal modes may be very important. Therefore, now we investigate the spectrum of small localized excitations of the DSG kink (antikink) using a standard method. Namely, we add a small perturbation $\delta\phi(x,t)$ to the static DSG kink $\phi_\mathrm{k}^{}(x)$:
\begin{equation}
\phi(x,t) = \phi_\mathrm{k}^{}(x) + \delta\phi(x,t), \quad |\delta\phi| \ll |\phi_\mathrm{k}^{}|.
\end{equation}
Substituting this $\phi(x,t)$ into the equation of motion \eqref{eq:eqmo}, and linearizing in $\delta\phi$, we obtain the partial differential equation for $\delta\phi(x,t)$:
\begin{equation}\label{eq:eq_for_delta_phi}
\frac{\partial^2\delta\phi}{\partial t^2} - \frac{\partial^2\delta\phi}{\partial x^2} + \left.\frac{d^2V}{d\phi^2}\right|_{\phi_\mathrm{k}^{}(x)}\cdot\delta\phi = 0.
\end{equation}
Looking for $\delta\phi$ in the form
\begin{equation}\label{eq:delta_phi}
\delta\phi(x,t) = \sum_n\eta_n^{}(x)\cos\:\omega_n^{} t,
\end{equation}
we obtain the Schr\"odinger-like eigenvalue problem
\begin{equation}\label{eq:Schrodinger}
\hat{H}\eta_n^{}(x) = \omega_n^2\eta_n^{}(x)
\end{equation}
with the operator $\hat{H}$ (the Hamiltonian)
\begin{equation}\label{eq:Schrod_Hamiltonian}
\hat{H} = -\frac{d^2}{dx^2} + U(x).
\end{equation}
Here the quantum-mechanical potential is
\begin{equation}\label{eq:Schrod_potential}
U(x) = \left.\frac{d^2V}{d\phi^2}\right|_{\phi_\mathrm{k}^{}(x)}.
\end{equation}
It can be easily shown that the discrete spectrum in the potential \eqref{eq:Schrod_potential} always possesses a zero mode $\omega_0^{}=0$. Differentiating eq.~\eqref{eq:eqmo_static} with respect to $x$, and taking into account that $\phi_\mathrm{k}^{}(x)$ is a solution of eq.~\eqref{eq:eqmo_static}, we see that
\begin{equation}
-\frac{d^2}{dx^2}\frac{d\phi_\mathrm{k}^{}}{dx} + \left.\frac{d^2V}{d\phi^2}\right|_{\phi_\mathrm{k}^{}(x)}\cdot\frac{d\phi_\mathrm{k}^{}}{dx} = 0,
\end{equation}
or, in other words,
\begin{equation}
\hat{H}\cdot\frac{d\phi_\mathrm{k}^{}}{dx} = 0\cdot\frac{d\phi_\mathrm{k}^{}}{dx}.
\end{equation}
So $\displaystyle\frac{d\phi_\mathrm{k}^{}}{dx}$ is really an eigenfunction of the Hamiltonian \eqref{eq:Schrod_Hamiltonian} associated with the zero frequency.

\begin{figure*}[h!]
\subfigure[The quantum-mechanical potential]{
\centering\includegraphics[width=0.45\textwidth]{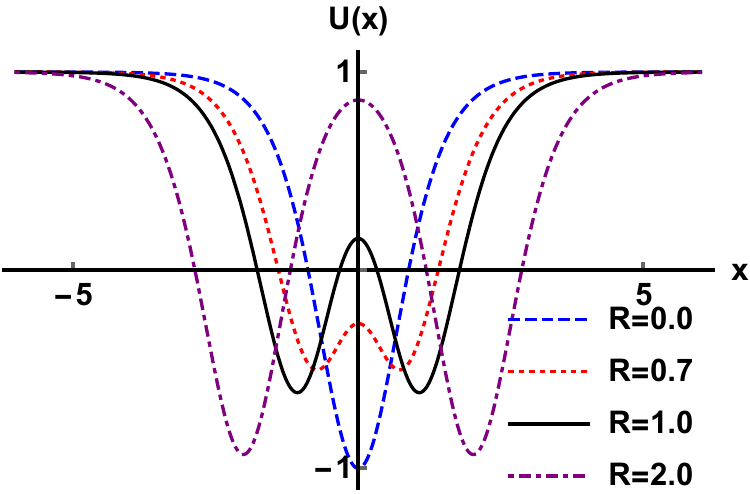}\label{fig:quantum-mechanical_potential} }
\subfigure[$R$-dependence of the vibrational mode of the kink]{
\centering\includegraphics[width=0.45\textwidth]{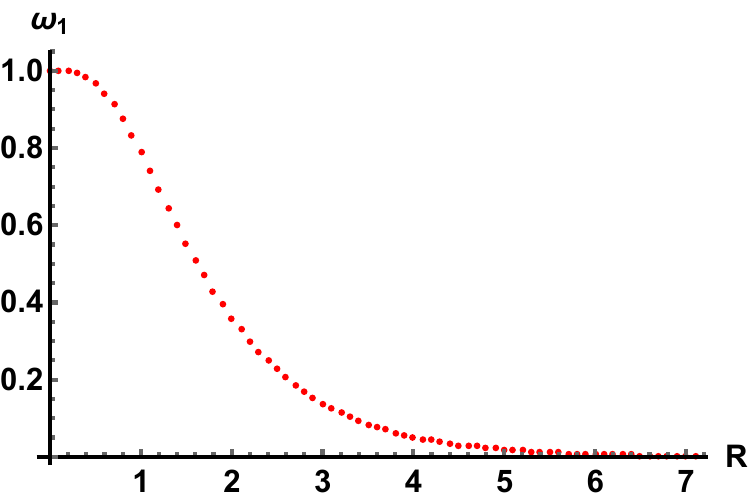}
\label{fig:Omega1R}}
\caption{(a) The quantum-mechanical potential \eqref{eq:Schrod_potential_DSG}, and (b) $R$-dependence of the vibrational mode's frequency.}
\end{figure*}

The potential $U(x)$ for the double sine-Gordon kink (antikink) can be obtained by substituting eqs.~\eqref{eq:potential_R} and \eqref{eq:DSG_kinks_1} in \eqref{eq:Schrod_potential},
\begin{equation*}
U(x) = \frac{8\tanh^2 R}{(1+\mbox{sech}^2R\sinh^2x)^2}
\end{equation*}
\begin{equation}\label{eq:Schrod_potential_DSG}
+\frac{2(3-4\cosh^2R)}{\cosh^2R}\frac{1}{1+\mbox{sech}^2R\sinh^2x}+1.
\end{equation}
The shape of the potential crucially depends on the parameter $R$, see fig.~\ref{fig:quantum-mechanical_potential}.
For $R=0$ equation \eqref{eq:Schrod_potential_DSG} gives the P\"oschl-Teller potential,
\begin{equation}\label{eq:Schrod_potential_DSG_zero}
U_0^{}(x) = 1-\frac{2}{\cosh^2x},
\end{equation}
which corresponds to the case of the sine-Gordon model. On the other hand, for $R\gg 1$ from eq.~\eqref{eq:Schrod_potential_DSG} we obtain
\begin{equation}\label{eq:Schrod_potential_DSG_infty}
U_\infty^{}(x) \approx \begin{cases}
1-\displaystyle\frac{2}{\cosh^2(x-R)}\ \mbox{for } ||x|-R|\lesssim 1,\\
\qquad\quad 1 \qquad\qquad\ \mbox{for } ||x|-R|\gg 1.
\end{cases}
\end{equation}

The discrete spectrum in the potential well \eqref{eq:Schrod_potential_DSG} for arbitrary value of $R$ can be obtained numerically by using a modification of the shooting method, see, {\it e.g.}, \cite{Gani.JHEP.2015,Belendryasova.arXiv.2017,Belendryasova.conf.2017}. First of all, for all $R$ there is the zero mode $\omega_0^{}=0$. Apart from that, we have found the vibrational mode $\omega_1^{}$ with the frequency that depends on $R$, see fig.~\ref{fig:Omega1R}.

At $R\to 0$ the frequency $\omega_1^{}$ goes to the boundary of the continuum, which corresponds to the sine-Gordon case. With increasing $R$ the frequency $\omega_1^{}$ decreases to zero. At large $R$'s the levels $\omega_0^{}$ and $\omega_1^{}$ are the result of splitting of the zero mode of each of the two potential wells \eqref{eq:Schrod_potential_DSG_infty}.

\section{Collisions of the double sine-Gordon kinks}
\label{sec:Scattering}

We studied the collision of the DSG kink and antikink. In order to do this, we used the initial configuration in the form of the DSG kink and the DSG antikink, centered at $x=-\xi$ and $x=\xi$, respectively, and moving towards each other with the initial velocities $v_\mathrm{in}^{}$. We solved the partial differential equation \eqref{eq:eqmo} with the $R$-parameterized potential \eqref{eq:potential_R} numerically, extracting the values of $\phi(x,0)$ and $\displaystyle\frac{\partial\phi(x,0)}{\partial t}$ from the following initial configuration:
\begin{equation*}
\phi(x,t) = \phi_\mathrm{k}^{}\left(\frac{x+\xi-v_\mathrm{in}^{}t}{\sqrt{1-v_\mathrm{in}^2}}\right) + \phi_\mathrm{\bar k}^{}\left(\frac{x-\xi+v_\mathrm{in}^{}t}{\sqrt{1-v_\mathrm{in}^2}}\right)-2\pi
\end{equation*}
\begin{equation*}
= 4\arctan \left[\frac{1}{\cosh R}\sinh \left(\frac{x+\xi-v_{in}^{}t}{\sqrt{1-v_\mathrm{in}^2}} \right)\right]
\end{equation*}
\begin{equation}\label{eq:incond}
- 4\arctan \left[\frac{1}{\cosh R}\sinh\left(\frac{x-\xi+v_{in}^{}t}{\sqrt{1-v_\mathrm{in}^2}}\right)\right]-2\pi.
\end{equation}

We discretized space and time using a grid with the spatial step $h$, and the time step $\tau$. We used the following discrete expressions for the second derivatives of the field:
\begin{equation}
\frac{\partial^2\phi}{\partial t^2} = \frac{11\phi_{n,j+1} -20\phi_{n,j}+6\phi_{n,j-1}+4\phi_{n,j-2}-\phi_{n,j-3}}{12\tau^2},
\end{equation}
\begin{equation}
\frac{\partial^2\phi}{\partial x^2} = \frac{-\phi_{n-2,j} +16\phi_{n-1,j}-30\phi_{n,j}+16\phi_{n+1,j}-\phi_{n+2,j}}{12h^2},
\end{equation}
where $\phi_{n,j}=\phi(nh,j\tau)$, $n=0,\pm 1,\pm 2,\dots$, and\\
$j=-3,-2,-1,0,1,2,\dots$.

We performed the numerical simulations for the steps $h=0.025$ and $\tau=0.005$, respectively, and for two different $\xi$: 10 and 20. We have also checked the stability of the results with respect to decrease of the steps. Fixed boundary conditions were used.

\begin{figure*}[t!]
\begin{center}
  \centering
  \subfigure[Kinks escape after one collision, $v_\mathrm{in}^{}=0.2500$]{\includegraphics[width=0.49\textwidth]{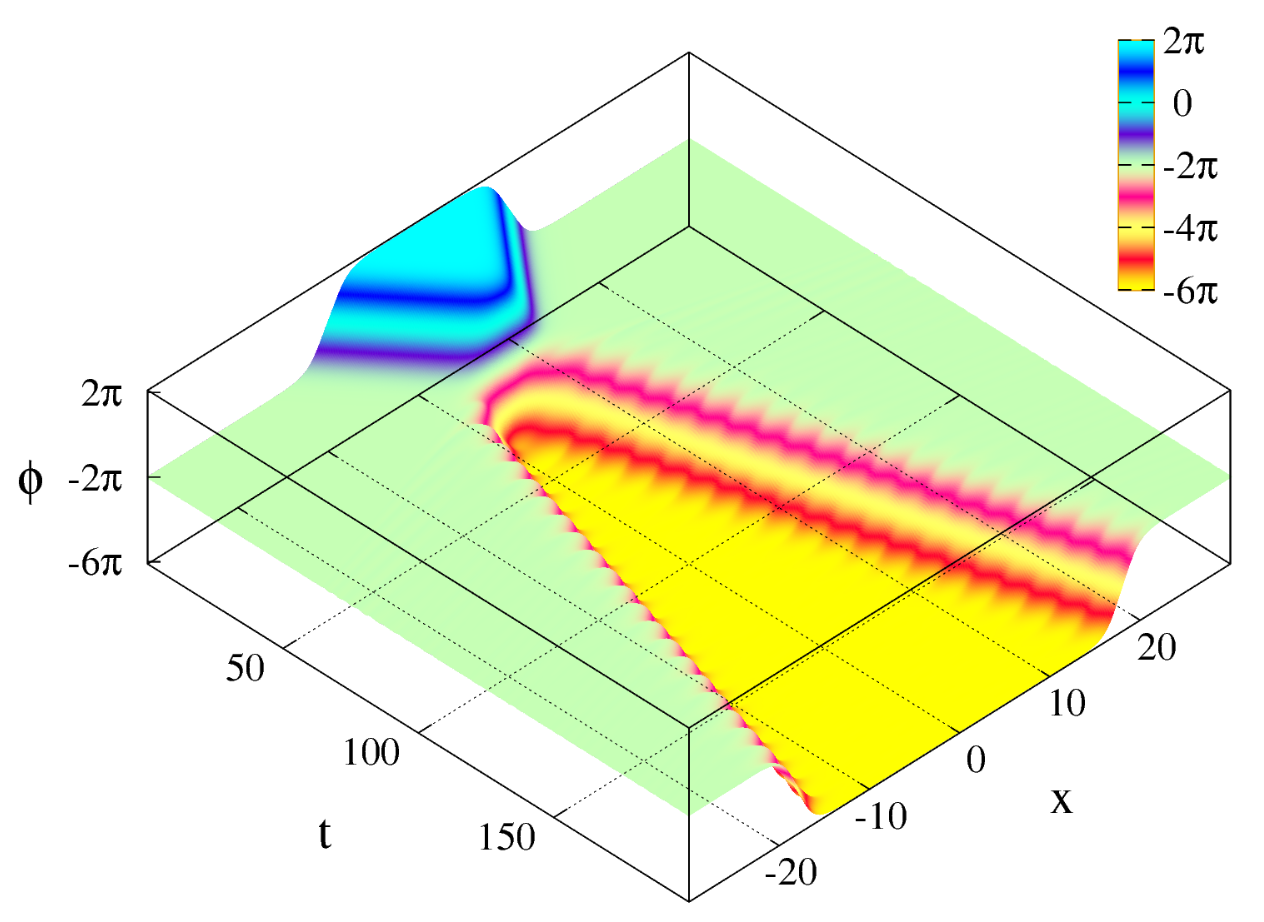}\label{fig:F3DR1V02500}}
  \subfigure[Bion formation, $v_\mathrm{in}^{}=0.2100$]
{\includegraphics[width=0.49\textwidth]{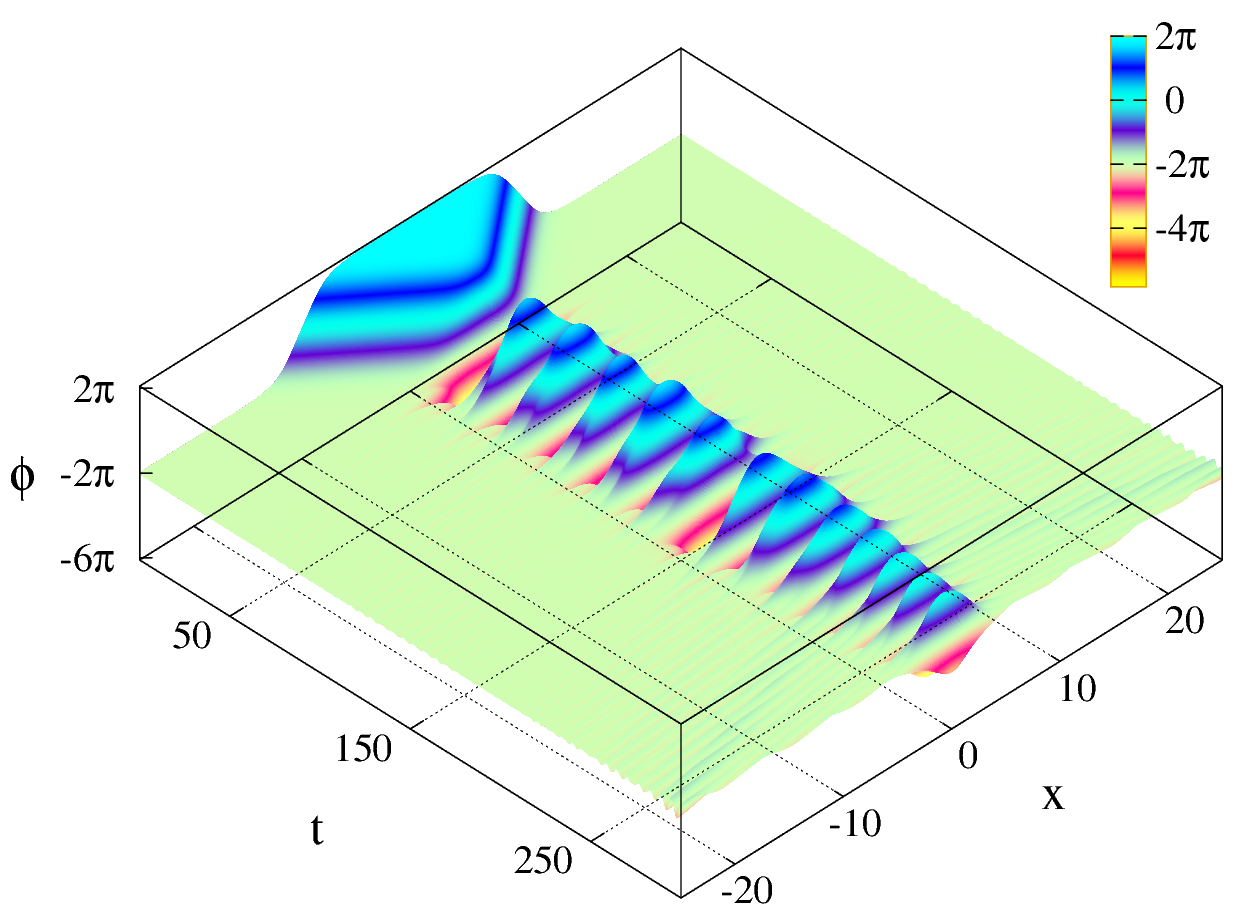}\label{fig:F3DR1V02100}}
  \\
 \subfigure[Two-bounce window, $v_\mathrm{in}^{}=0.2340$]{\includegraphics[width=0.49\textwidth]{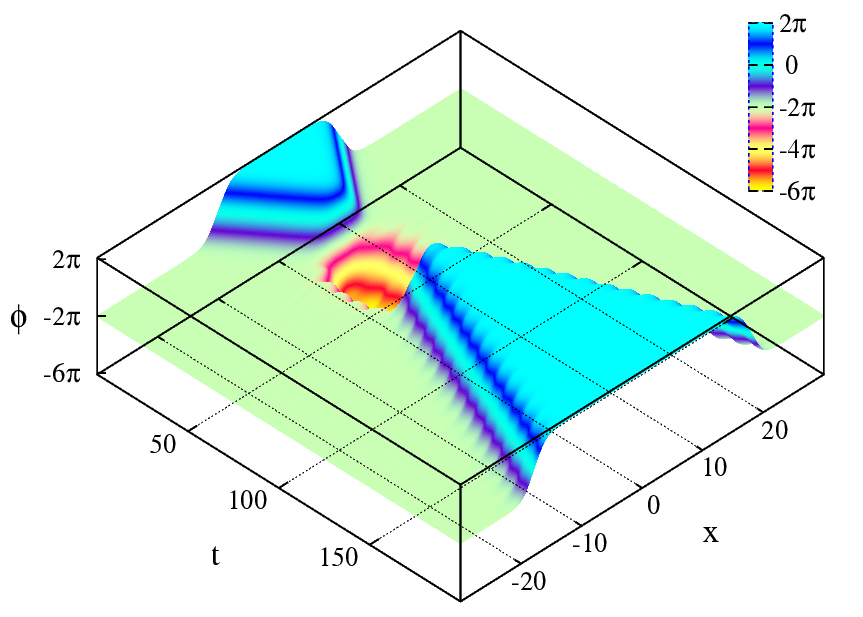}\label{fig:F3DR1V02340}}
  \subfigure[Three-bounce window, $v_\mathrm{in}^{}=0.2380$]
{\includegraphics[width=0.49\textwidth]{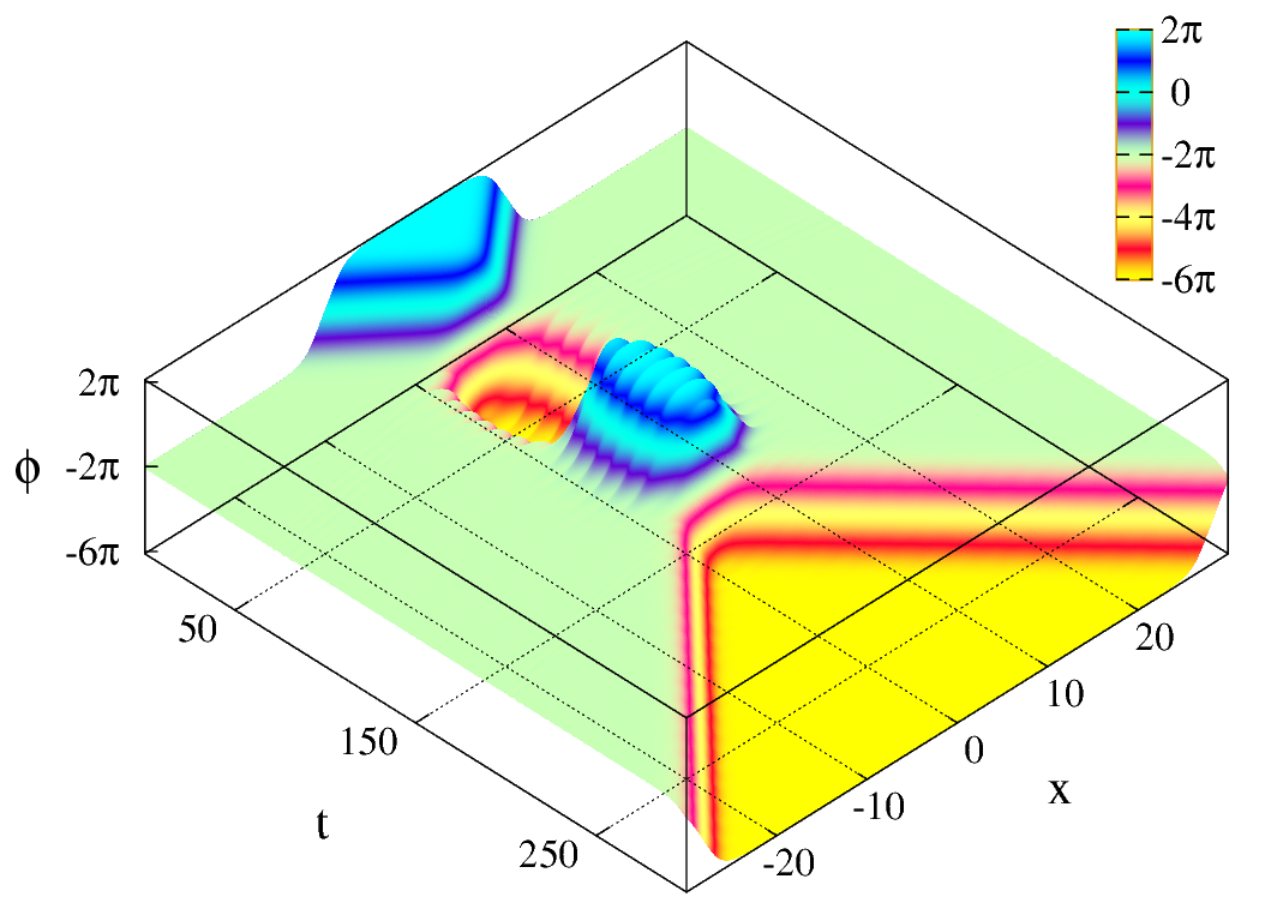}\label{fig:F3DR1V02380}}
  \caption{The space-time picture of the kink-antikink collisions at various initial velocities for $R=1$. (The initial half-separation is $\xi=10$ in these simulations.)}
  \label{}
\end{center}
\end{figure*}

In the kink-antikink collisions there is a critical value of the initial velocity, $v_\mathrm{cr}^{}$, which separates two different regimes of the collisions. At the initial velocities above the critical value, $v_\mathrm{in}^{}>v_\mathrm{cr}^{}$, the DSG kinks pass through each other and escape to infinities after one collision, see fig.~\ref{fig:F3DR1V02500}.
At $v_\mathrm{in}^{}<v_\mathrm{cr}^{}$ one observes the kinks' capture and formation of their long-living bound state --- a bion, see fig.~\ref{fig:F3DR1V02100}. At the same time, in the range $v_\mathrm{in}^{}<v_\mathrm{cr}^{}$ the so-called ``escape windows'' have been found. An escape window is a narrow interval of the initial velocities, at which the kink and the antikink escape after two, three, or more collisions, see figs.~\ref{fig:F3DR1V02340}, \ref{fig:F3DR1V02380}.

\subsection{The $R$-dependence of the critical velocity}

\begin{figure}[h!]
\centering
\includegraphics[width=0.55\textwidth]{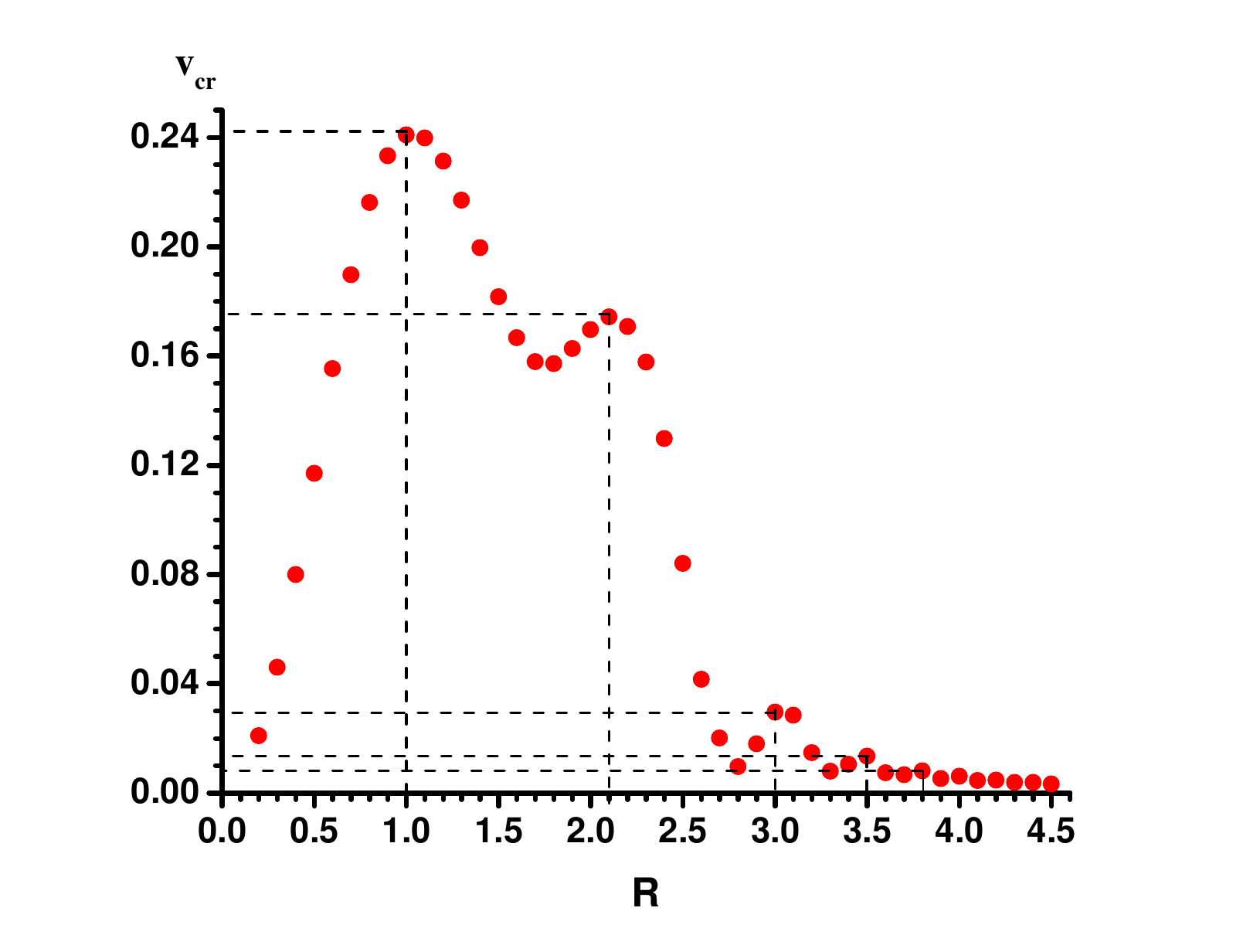} 
\caption{The critical velocity $v_\mathrm{cr}^{}$ as a function of the parameter $R$. (The initial half-separation is $\xi=20$ in these calculations.)} 
\label{fig:critical_velocity_vs_R}
\end{figure}

First of all, we found the dependence of the critical velocity $v_\mathrm{cr}^{}$ on the parameter $R$. Our results are shown in fig.~\ref{fig:critical_velocity_vs_R}.
One can see a series of peaks on the curve $v_\mathrm{cr}^{}(R)$, see also table \ref{tab:Table1}.
\begin{table}
\caption{Positions of the local maxima of the dependence $v_\mathrm{cr}^{}(R)$, which is shown in fig.~\ref{fig:critical_velocity_vs_R}.}
\label{tab:Table1}
\begin{center}
\begin{tabular}{|c|c|c|}
\hline
$n$ & $R_n^\mathrm{(max)}$ & $v_\mathrm{cr}^{}(R_n^\mathrm{(max)})$ \\
\hline
1 & 1.0 & 0.2409 \\
2 & 2.1 & 0.1743 \\
3 & 3.0 & 0.0296 \\
4 & 3.5 & 0.0135 \\
5 & 3.8 & 0.0081 \\
6 & 4.0 & 0.0061 \\
7 & 4.2 & 0.0047 \\
\hline
\end{tabular}
\end{center}
\end{table}
Note that at this point we have some discrepancy with the results of \cite{Campbell.dsG.1986}. The authors of \cite{Campbell.dsG.1986} report only one maximum of the dependence $v_\mathrm{cr}^{}(R)$ at $R\approx 1$. Probably it can be a consequence of small amount of experimental points in \cite{Campbell.dsG.1986}.

From fig.~\ref{fig:critical_velocity_vs_R} one can see that the critical velocity turns to zero at $R=0$, which corresponds to the integrable sine-Gordon model, see eq.~\eqref{eq:potential_R_limits}. Besides that, $v_\mathrm{cr}^{}$ decreases to zero with increasing $R$ at large $R$. Remind here, that the limit $R\to+\infty$ also corresponds to the case of the integrable sine-Gordon model, as one can see from eq.~\eqref{eq:potential_R_limits}. Therefore it is quite natural that the critical velocity has a maximum at some $R$ and tends to zero at $R\to 0$ and $R\to+\infty$. The presence of a {\it series} of local maxima on the curve is an interesting fact that is observed for the first time. Apparently we can assume that one of the maxima is the main (probably, $R_1^\mathrm{(max)}$), while the other ones appear due to some change of the kink-antikink interaction in the collision process with increasing $R$. At large values of $R$ the DSG kink splits into two sine-Gordon solitons. Therefore, we can assume that in the DSG kink-antikink collision the four sine-Gordon solitons interact pairwise. This transition from the simple collision of the DSG kinks to more complicated pairwise interaction of the sine-Gordon solitons can lead, in particular, to the non-monotonicity of the dependence $v_\mathrm{cr}^{}(R)$ at $R>R_1^\mathrm{(max)}$.

\subsection{Two oscillons in the final state}

\begin{figure*}[t!]
\begin{center}
  \centering
  \subfigure[$v_\mathrm{in}^{}=0.1847$, $v_\mathrm{f}^{}\approx 0.10$]{\includegraphics[width=0.49\textwidth]{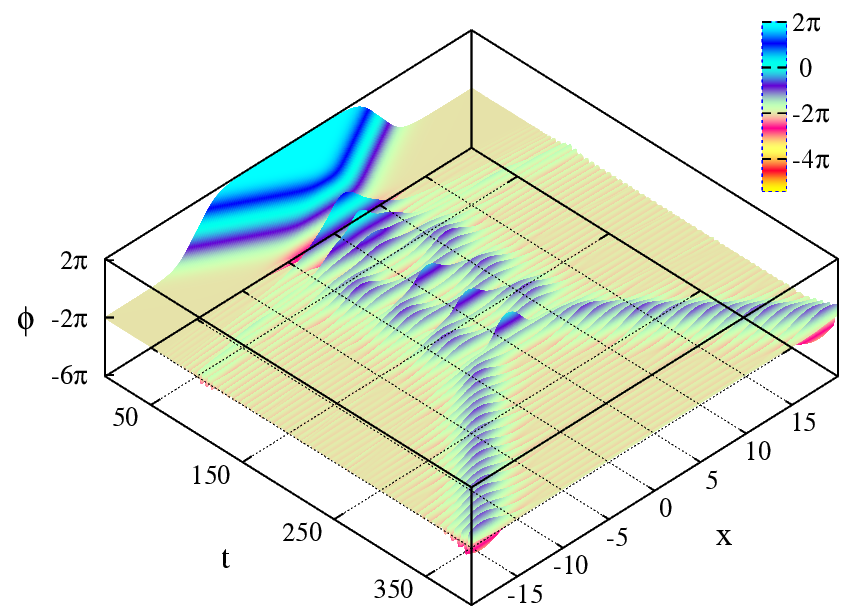}\label{fig:F3DR1V01847}}
  \subfigure[$v_\mathrm{in}^{}=0.18467$, $v_\mathrm{f}^{}\approx 0.03$]
{\includegraphics[width=0.49\textwidth]{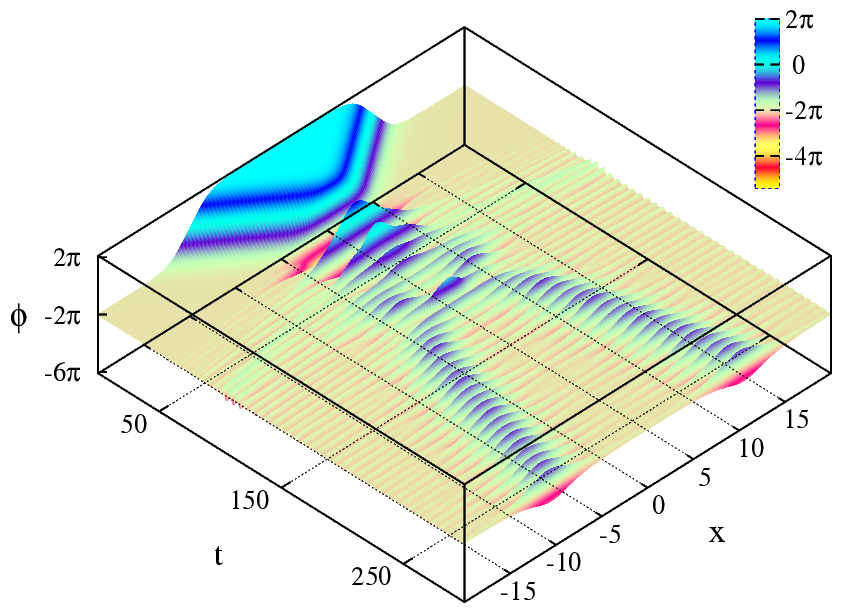}\label{fig:F3DR1V018467}}
  \subfigure[$v_\mathrm{in}^{}=0.18470001$, $v_\mathrm{f}^{}\approx 0.15$]
{\includegraphics[width=0.49\textwidth]{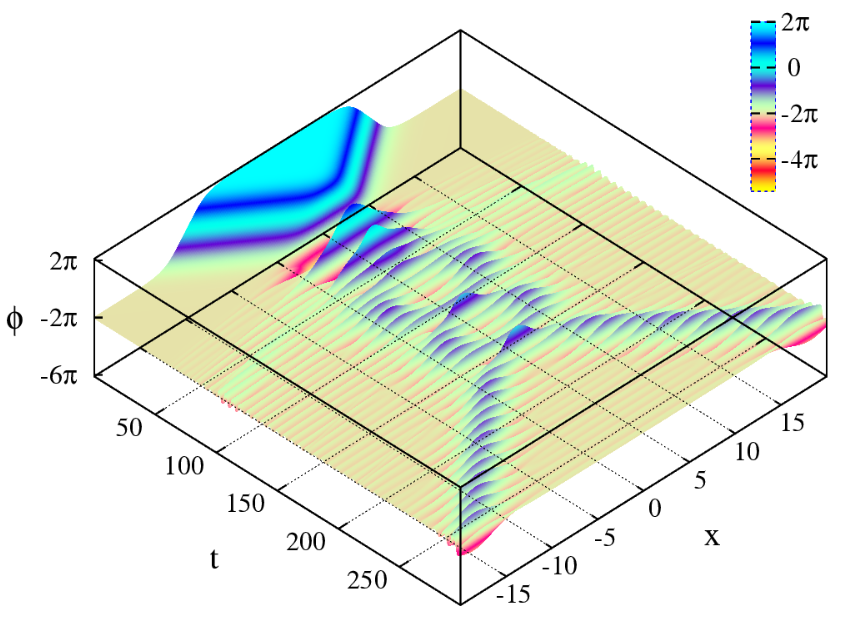}\label{fig:F3DR1V018470001}}
 \subfigure[$v_\mathrm{in}^{}=0.18470003$, $v_\mathrm{f}^{}\approx 0.19$]{\includegraphics[width=0.49\textwidth]{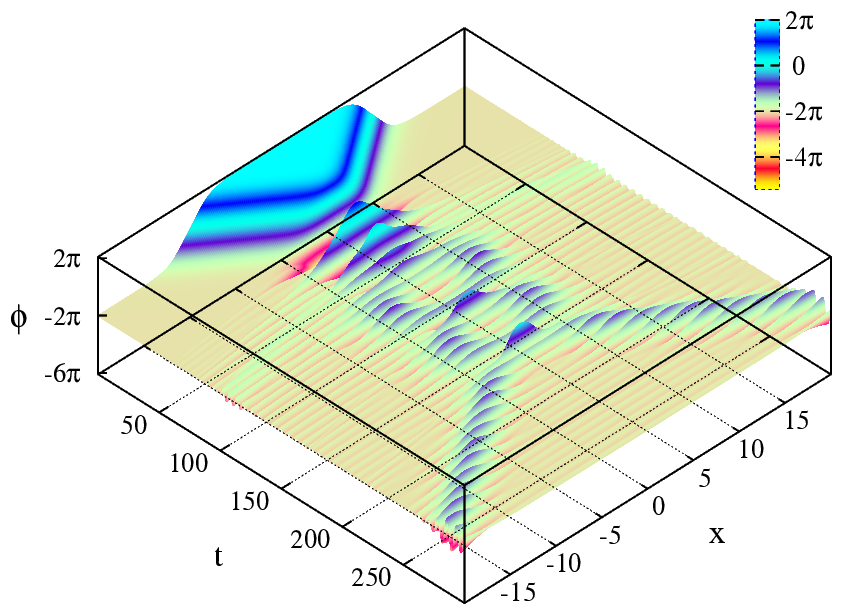}\label{fig:F3DR1V018470003}}
    \subfigure[$v_\mathrm{in}^{}=0.18472$, $v_\mathrm{f}^{}=0.00$]{\includegraphics[width=0.49\textwidth]{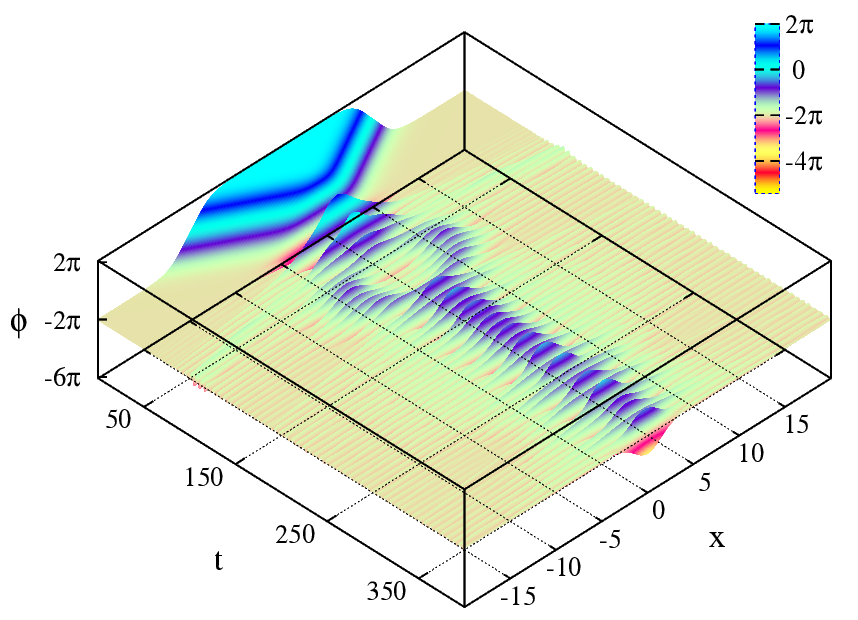}\label{fig:F3DR1V018472}}
  \subfigure[$v_\mathrm{in}^{}=0.18473$, $v_\mathrm{f}^{}\approx 0.07$]
{\includegraphics[width=0.49\textwidth]{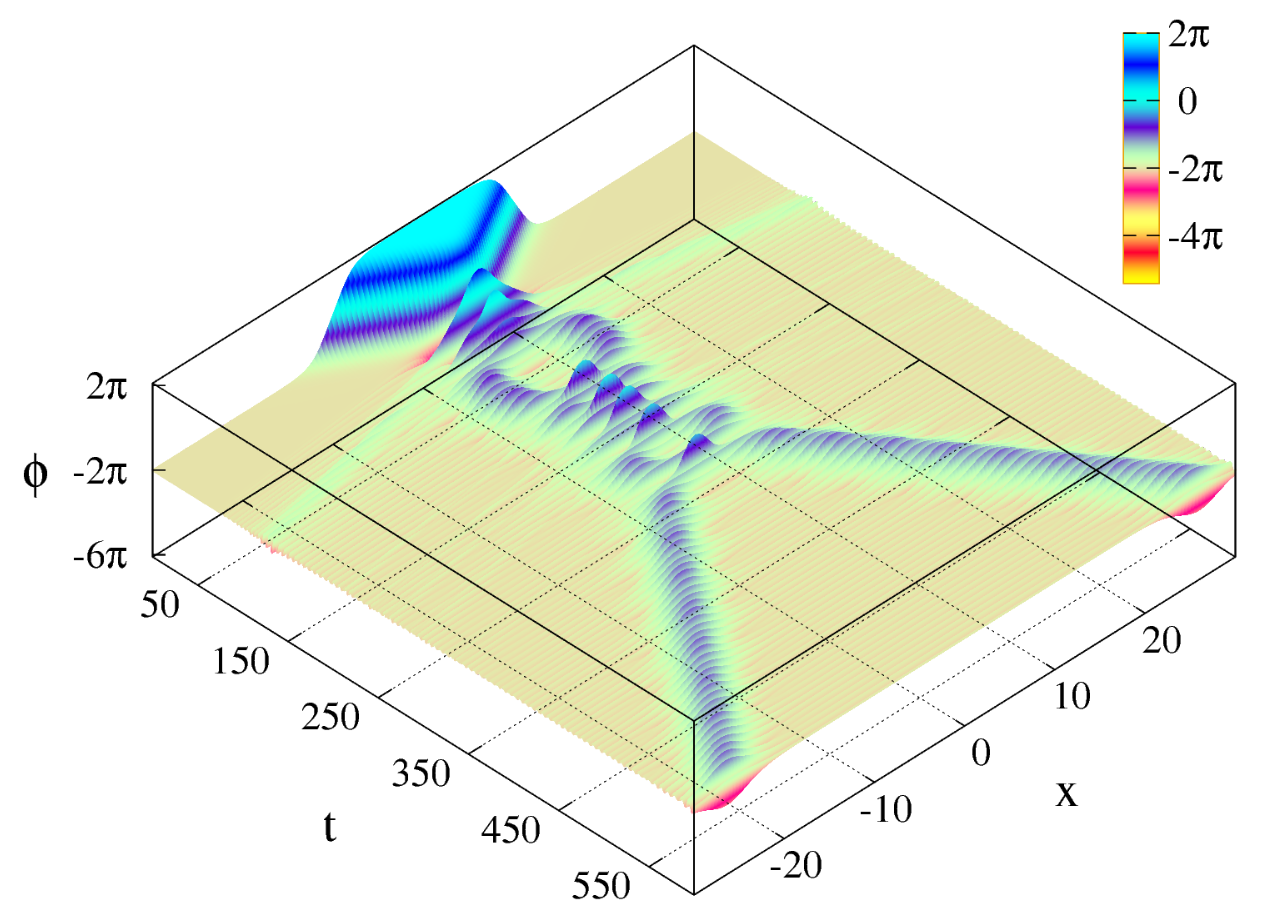}\label{fig:F3DR1V018473}}
  \caption{Kink-antikink collisions at various initial velocities ($R=1$, $\xi=10$) and escape of oscillons.}
  \label{fig:two_Breathers_in_the_final_state}
\end{center}
\end{figure*}

In the kink-antikink collisions below the critical velocity we observed a phenomenon, which, to the best of our knowledge, has not been reported for the double sine-Gordon model before. At some initial velocities of the colliding kinks we observed final configuration in the form of two escaping oscillons. At the same time, at some initial velocities we found formation of the configuration, which we can classify as a bound state of two oscillons. In fig.~\ref{fig:two_Breathers_in_the_final_state} we show some typical scenarios of that kind.

For example, at the initial velocity $v_\mathrm{in}^{}=0.1847$ we observe formation of a bound state of the kink and the antikink (a bion), which then evolves into two oscillons. These two oscillons are moving from each other, then stop, and start moving back to the collision point. This repeats several times, and after that the oscillons escape to infinities with the final velocity $v_\mathrm{f}^{}\approx 0.10$,
see fig.~\ref{fig:F3DR1V01847}.

Formation of the bound state of oscillons and the escape of oscillons are extremely sensitive to changes of the initial velocity of the colliding kinks. For example, at the initial velocity $v_\mathrm{in}^{}=0.18467$ the oscillons escape to infinities after fewer number of collisions, see fig.~\ref{fig:F3DR1V018467}. The final velocity of the escaping oscillons also differs substantially: at $v_\mathrm{in}^{}=0.1847$ we obtain $v_\mathrm{f}^{}\approx 0.10$
(fig.~\ref{fig:F3DR1V01847}), while at $v_\mathrm{in}^{}=0.18467$ we have $v_\mathrm{f}^{}\approx 0.03$
(fig.~\ref{fig:F3DR1V018467}), and at $v_\mathrm{in}^{}=0.18470001$ we obtain $v_\mathrm{f}^{}\approx 0.15$
(fig.~\ref{fig:F3DR1V018470001}). At the initial velocity of the colliding kinks $v_\mathrm{in}^{}=0.18470003$, fig.~\ref{fig:F3DR1V018470003}, the final velocity of the escaping oscillons is $v_\mathrm{f}^{}\approx 0.19$.

In the kink-antikink collision at the initial velocity $v_\mathrm{in}^{}=0.18472$, fig.~\ref{fig:F3DR1V018472}, we observe formation of two oscillons, which are moving apart from each other, then approach and collide. After that we have final configuration in the form of the oscillating configuration of the type of bion at the origin. Apparently this final configuration can be viewed as a bound state of two oscillons, which oscillate around each other with small amplitude.

At the initial velocity $v_\mathrm{in}^{}=0.18473$, fig.~\ref{fig:F3DR1V018473}, we observe even more complicated picture. The kinks collide, form a bion, which, in turn, decays into two oscillons. These oscillons escape at some distance and then collide again. After that, for some time we observe the bound state of oscillons --- small amplitude oscillations of oscillons around each other. Finally, the oscillons escape at some valuable distance, collide for the last time, and escape to infinities with the final velocities $v_\mathrm{f}^{}\approx 0.07$.

The obtained results show that in the DSG kink-antikink scattering we found new phenomenon --- formation of the pair of oscillons, which can form a bound state or escape to spatial infinities. Note that similar behaviour has been observed recently in the collisions of kinks of another model with non-polynomial potential \cite{Bazeia.arXiv.2017.sinh,Bazeia.arXiv.2017.sinh.conf}.

\section{Conclusion}
\label{sec:Conclusion}

We have studied the scattering of kinks of the double sine-Gordon model. Several different parameterizations of this model are known in the literature. We used the so-called $R$-parameterization, in which the potential of the model depends on the positive parameter $R$, see eq.~\eqref{eq:potential_R}.

The scattering of the DSG kink and antikink looks as follows. There is a critical value of the initial velocity $v_\mathrm{cr}$ such that at $v_\mathrm{in}>v_\mathrm{cr}$ the kinks pass through each other and then escape to infinities. At $v_\mathrm{in}<v_\mathrm{cr}$ one observes formation of a bound state of the kinks --- a bion. Besides that, at some narrow intervals of the initial velocity (which are called ``escape windows'') from the range $v_\mathrm{in}<v_\mathrm{cr}$ the kinks escape to infinities after two or more collisions.

We have obtained the dependence of the critical velocity $v_\mathrm{cr}$ on the parameter $R$. The curve $v_\mathrm{cr}(R)$ has several well-seen local maxima, see fig.~\ref{fig:critical_velocity_vs_R} and table \ref{tab:Table1}. Note some discrepancy between our results and the results of \cite{Campbell.dsG.1986}. The authors of \cite{Campbell.dsG.1986} reported only one maximum of the curve $v_\mathrm{cr}(R)$. This could be a consequence of small number of experimental points between $R=1.8$ and $R=2.4$ presented in \cite{Campbell.dsG.1986}.

Apart from the previously known bions and escape windows, in the range $v_\mathrm{in}<v_\mathrm{cr}$ in our numerical experiments we observed a new phenomenon, which could be classified as formation of a bound state of two oscillons, and their escape in some cases. So at some initial velocities of the colliding kinks, in the final state we observed two oscillons escaping from the collision point. The time between the first kinks impact and the beginning of the oscillons escaping can be rather big. The field evolution during this time is quite complicated. First, we observe formation of a bion. After a short time, this bion evolves into a configuration, which can be identified as a bound state of two oscillons oscillating around each other. The amplitude of these oscillations can vary substantially. After that the oscillons either remain bound or escape to spatial infinities, depending on the initial velocity of the colliding kinks. It is interesting that formation of a bound state of two oscillons, as well as escape of oscillons, has been found recently in the collisions of kinks of the sinh-deformed $\varphi^4$ model \cite{Bazeia.arXiv.2017.sinh,Bazeia.arXiv.2017.sinh.conf}. We think that this new phenomenon can be a part of new interesting physics within a wide class of non-linear models.

We can assume that the escape of oscillons is a kind of resonance phenomena, {\it i.e.} it is a consequence of the resonant energy exchange between oscillon's kinetic energy and its internal vibrational degree(s) of freedom. A detailed study of such exchange could be a subject of future work.

In conclusion, we would like to mention several issues that we think could become a subject of future study.
\begin{itemize}
\item
First, it would be interesting to explain the behaviour of the dependence $v_\mathrm{cr}(R)$ with a series of local maxima. This non-monotonicity could be a consequence of the kink's shape changing with increasing of the parameter $R$. So at large $R$'s the interaction of the DSG kinks could be reduced to pairwise interaction of the subkinks, which are the sine-Gordon solitons separated by the distance $2R$. Note that the authors of \cite{Demirkaya.cc.2017} observed the non-monotonic dependence of $v_\mathrm{cr}$ on the model parameter in the parametrically modified $\varphi^6$ model. In order to explain the phenomenon, they applied the collective coordinate approach. We believe that similar analysis could be applied to the double sine-Gordon kink-antikink system.
\item
Second, the oscillons escape in the final state, as well as formation of a bound state of two oscillons, are new interesting phenomena, which have to be explained qualitatively and probably quantitatively.
\item
Third, it would be very interesting to study multikink collisions within the DSG model in the spirit of \cite{Moradi.JHEP.2017}. Due to complex internal structure of the DSG kinks, the multikink collisions could result in a rich variety of new phenomena.
\end{itemize}
Answers to these questions would substantially improve our understanding of the DSG kinks dynamics.

\section*{Acknowledgments}

This research was supported by the MEPhI Academic Excellence Project (contract No.\ 02.a03.21.0005, 27.08.2013).

\end{document}